\documentclass[prl,aps,amssymb,superscriptaddress,twocolumn, hypertext]{revtex4-2} 
\usepackage{hyperref}
\hypersetup{
    colorlinks,
    linkcolor={black},
    citecolor={blue!80!black},
    urlcolor={blue!80!black}
}

\usepackage{graphicx}
\usepackage{dcolumn}
\usepackage{bm}
\usepackage{soul}
\usepackage[utf8]{inputenc}
\usepackage{amsmath}
\usepackage{multirow}
\usepackage{xcolor}
\usepackage[per-mode=symbol]{siunitx}
\usepackage{braket}
\usepackage{textcomp}
\DeclareSIUnit\gauss{G}
\graphicspath{{/}}

\makeatletter
\renewcommand*{\@fnsymbol}[1]{\ensuremath{\ifcase#1\or \dagger \or *\or \ddagger\or
   \mathsection\or \mathparagraph\or \|\or **\or \dagger\dagger
   \or \ddagger\ddagger \else\@ctrerr\fi}}
\makeatother

\begin{document}


\title{Retrieving Lost Atomic Information: Monte Carlo-based Parameter Reconstruction of an Optical Quantum System}

\author{Laura Orphal-Kobin}\thanks{L.O.-K. and G.P. contributed equally to this work.}
\affiliation{\vspace{0.5em}Department of Physics \& IRIS Adlershof, Humboldt-Universität zu Berlin, Newtonstr. 15, 12489 Berlin, Germany}
\author{Gregor Pieplow}\thanks{L.O.-K. and G.P. contributed equally to this work.}
\affiliation{\vspace{0.5em}Department of Physics \& IRIS Adlershof, Humboldt-Universität zu Berlin, Newtonstr. 15, 12489 Berlin, Germany}
\author{Alok Gokhale} 
\affiliation{\vspace{0.5em}Department of Physics \& IRIS Adlershof, Humboldt-Universität zu Berlin, Newtonstr. 15, 12489 Berlin, Germany}
\author{Kilian Unterguggenberger} 
\affiliation{\vspace{0.5em}Department of Physics \& IRIS Adlershof, Humboldt-Universität zu Berlin, Newtonstr. 15, 12489 Berlin, Germany}
\author{Tim Schröder}
\email{tim.schroeder@physik.hu-berlin.de}
\affiliation{\vspace{0.5em}Department of Physics \& IRIS Adlershof, Humboldt-Universität zu Berlin, Newtonstr. 15, 12489 Berlin, Germany}
\affiliation{\vspace{0.5em}Ferdinand-Braun-Institut, Gustav-Kirchhoff-Str. 4, 12489 Berlin, Germany}

\date{\today}
\begin{abstract}
\noindent 
In regimes of low signal strengths and therefore a small signal-to-noise ratio, standard data analysis methods often fail to accurately estimate system properties. We present a method based on Monte Carlo simulations to effectively restore robust parameter estimates from large sets of undersampled data. This approach is illustrated through the analysis of photoluminescence excitation spectroscopy data for optical linewidth characterization of a nitrogen-vacancy color center in diamond. We evaluate the quality of parameter prediction using standard statistical data analysis methods, such as the median and the Monte Carlo method. Depending on the signal strength, we find that the median can be precise (narrow confidence intervals) but very inaccurate. A detailed analysis across a broad range of parameters allows to identify the experimental conditions under which the median provides a reliable predictor of the quantum emitter's linewidth. We also explore machine learning to perform the same task, forming a promising addition to the parameter estimation toolkit. Finally, the developed method offers a broadly applicable tool for accurate parameter prediction from low signal data, opening new experimental regimes previously deemed inaccessible.
\end{abstract}

\maketitle

%
\begin{figure*}[ht!]
\centering
\includegraphics[width=0.9\textwidth]{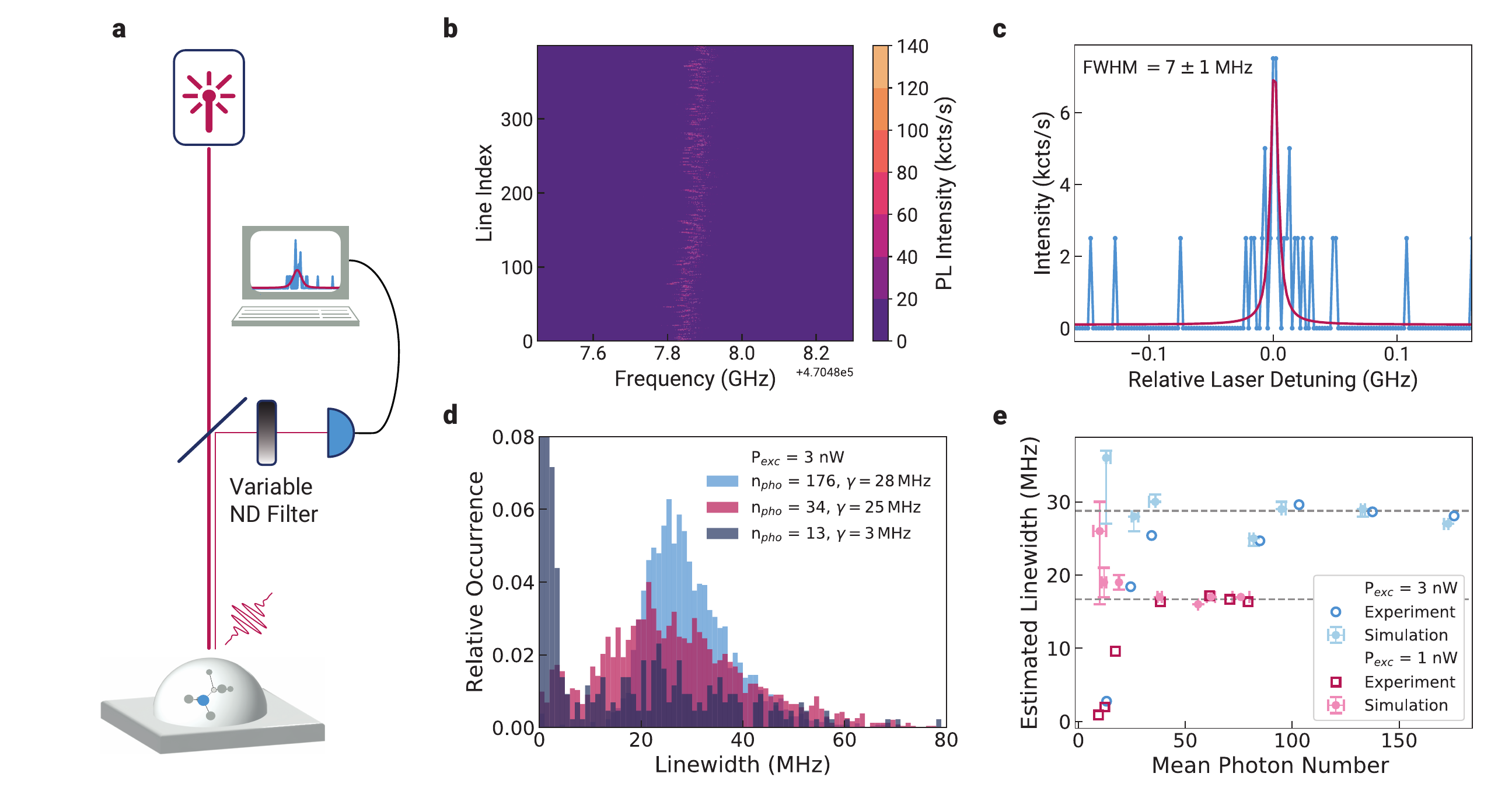} 
\caption{\label{fig:exp}(a) Experiment: In a confocal microscope the number of detected photons is controlled by a variable neutral density filter. (b) PLE scans at an excitation power of 3\,nW.
(c) FWHM extracted from individual line scans. For low signal strenghts, standard fitting routines fail to extract the physical linewidth. (d) Histograms summarizing the relative occurrences of linewidths for different photon collection efficiencies at 3\,nW excitation power showing that low photon collection efficiencies, produce nonphysical linewidths. (e) Estimated experimental ``single-scan'' linewidth using the median as a function of detected photon number for excitation powers of 3\,nW (blue circles) and 1\,nW (red squares). The light blue and light red markers with errorbars (99$\%$ confidence intervals) indicate linewidths extracted from Monte Carlo simulations. The true linewidths of 17\,MHz and 29\,MHz for excitation powers of 1\,nW and 3\,nW, respectively, correspond to the dashed line (averaged median of the three last points).}
\end{figure*}
Random sampling techniques are powerful tools to model the behavior of complex systems, approximating solutions, and predicting outcomes in situations where deterministic solutions are challenging to obtain. Advanced statistical methods, such as Monte Carlo parameter estimation, are commonly used in astro- and particle physics~\cite{bohm_comparison_2012,gagunashvili_parametric_2011,bohm_introduction_2017,zech_comparing_1995}. Here, these techniques are employed to address the challenges posed by limited photon detection for the optical characterization of quantum emitters. Moreover, machine learning is explored for parameter prediction~\cite{borodinov_deep_2019, nguyen_deep_2021, nolan_machine_2021, xiao_parameter_2022, yang_entanglement_2024}. 

The precise optical characterization of atom-like systems is crucial for addressing fundamental questions, and for assessing their potential in emerging quantum technologies. Artificial atoms are a resource in secure communication~\cite{basset_quantum_2021,ekert_ultimate_2014}, quantum information processing~\cite{wehner_quantum_2018}, and, moreover, in biology and medicine~\cite{azam_carbon_2021,wagner_quantum_2019,wu_diamond_2016}. Already established platforms for quantum emitters in solid-state systems~\cite{atature_material_2018,wolfowicz_quantum_2021} are color centers in diamond~\cite{orphal-kobin_coherent_nodate}, silicon carbide~\cite{castelletto_silicon_2020}, quantum dots~\cite{vajner_quantum_2022}, defects in two-dimensional materials~\cite{kianinia_quantum_2022}, and rare-earth ions~\cite{zhong_emerging_2019,pettit_perspective_2023}. 
Key parameters for the characterization of these systems include the homogeneous and inhomogeneous linewidth of optically excited states~\cite{yurgens_spectrally_2022,chakravarthi_impact_2021,ruf_optically_2019,van_dam_optical_2019,van_de_stolpe_check-probe_2024,pedersen_near_2020,liu_single_2018}, the Hong-Ou-Mandel (HOM) Visibility of emitted photons~\cite{Hong1987}, as well as spin lifetime and spin coherence time~\cite{muhonen_storing_2014,rosenthal_microwave_2023}. Particularly, the control of the optical emission resonance is a prerequisite for photon-mediated entanglement protocols~\cite{kambs_limitations_2018,beukers_remote-entanglement_2024}. Depending on the system under investigation, the characterization of isolated quantum emitters can become time- and resource-intensive due to low emission efficiencies caused by non-radiative processes~\cite{gali_ab_2019}, internal reflection within materials with high refractive indices~\cite{SchroederJOSAB2016}, and low photon collection efficiencies in experimental setups.
A well-known example of a system that encounters these challenges is the negatively charged nitrogen-vacancy color center in diamond (NV). While the NV is very well characterized, individual NVs often exhibit poor optical performance due to a comparatively small Debye-Waller factor~\cite{yurgens_cavity-assisted_2024,LiNatComms2015} and pronounced spectral diffusion~\cite{orphal-kobin_optically_2023,WoltersPRL2013}. Mitigating these limitations could significantly enhance their utility in characterizing photonic integrated circuits~\cite{wan_large-scale_2020,SipahigilScience2016,riedel_efficient_2023}, including the rapid prototyping of novel nanostructures~\cite{Bopp2022}. 

In this work, advanced statistical methods are employed as a technique to restore information from undersampled data. For instance, full information on the linewidth of an NV is obtained from photoluminescence excitation spectroscopy (PLE) data even with minimal detection signals.  
This study explores the limits of traditional data analysis methods in PLE experiments with low photon collection efficiencies. Using synthetic data, the quality of parameter prediction by the median and the Monte Carlo method (MCM) is analyzed in a broad parameter regime.
The presented method can be transferred to any physical problem where data are obscured by low signal transmission and statistical noise. We share our Monte Carlo analysis code and example data to support researchers across various fields~\cite{mc_github}.


\begin{figure}
\centering
\includegraphics[width= .75\columnwidth]{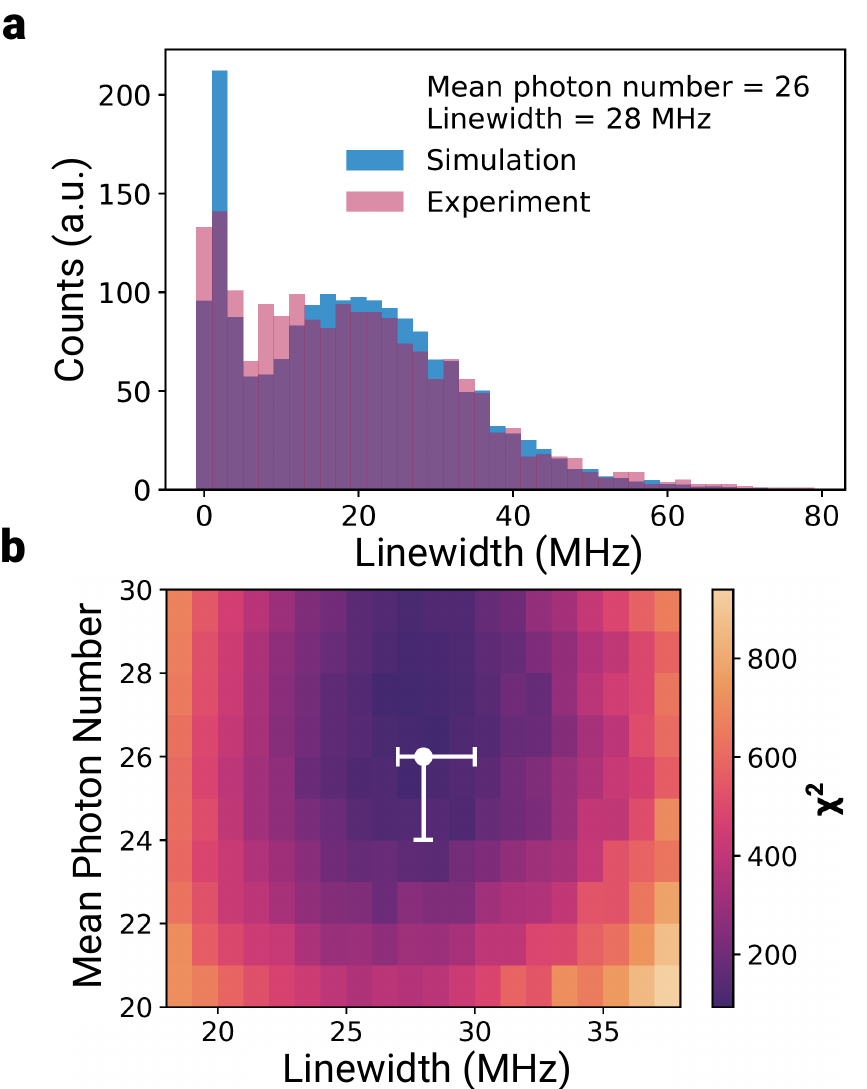}
\caption{\label{fig:MC_method} (a) The MCM simulates PLE spectra and the expected linewidth distributions depending on the linewidth of the emitter and the collected photon number. The histograms are fitted to the experimental results. (b) Using a $\chi^2$-test, the best estimate together with the $99\%$ confidence intervals is determined.}
\end{figure}

In a well controlled PLE experiment on a single NV center in diamond (see Supplemental Material for experimental details), we deliberately enter a regime of noisy data 
where conventional statistical data analysis methods begin to fail. In this regime a more sophisticated method based on Monte Carlo simulations needs to be applied to obtain a reliable ``single-scan'' linewidth. 
We control the number of photons reaching the detector by incorporating a variable neutral density filter in the detection path, as illustrated in Fig.~\ref{fig:exp}a.
For every experimental setting a large set of 3200 PLE scans is recorded.  
In Fig.~\ref{fig:exp}b, a set of 400 scans is shown as an example. Individual PLE scans are fitted by a Voigt profile to determine the full width at half maximum (FWHM). In a regime of extremely low signals, the fitting routine ``fails'' leading to either illusory narrow linewidths extracted with high precision, as shown in Fig.~\ref{fig:exp}c, or overestimated linewidths with large uncertainty, discussed in Supplemental Material. These errors arise because statistical fluctuations in photon detection, known as shot noise, induce an uncertainty proportional to $\sqrt{n}$, where $n \in \{0,5\}$ is the number of detected photons per bin. In such cases, the limited photon count leads to significant deviations in linewidth fitting.
In Figure~\ref{fig:exp}d, the distribution of ``single-scan'' linewidths at an excitation power of 3\,nW is plotted for various collected photon numbers achieved by adjusting the transmission settings in the detection path. Notably, at extremely low photon numbers, the distribution shifts toward unphysically narrow linewidths. 

The ``single-scan'' linewidth is estimated from the median value of the distribution of FWHMs. Among the standard linewidth estimation techniques, we identify the median to be a robust predictor in a low photon number regime.
The linewidth is investigated as a function of the mean photon number collected in a frequency window of 150\,MHz around the resonance (see Fig.~\ref{fig:exp}e). Even though the physical linewidth of the NV emitter remains unchanged, for lower detected photon numbers a decrease in the estimated linewidth below the lifetime-limit is observed. For enhanced photon collection efficiencies, the estimated linewidth converges to the true ``single-scan'' linewidth and in agreement with a lifetime measurement determining a natural linewidth of 15\,MHz. Notably, the linewidth estimated from simulations (introduced below) only deviates for the lowest photon count by more than $10\%$, however, the confidence intervals still encompass the true value. By increasing the excitation power, the homogeneous absorption linewidth of the NV is affected by power broadening~\cite{noauthor_atomphoton_1998}.
While for the lower excitation power of 1\,nW the estimated linewidth is 17\,MHz and close to the lifetime-limited linewidth, at an excitation power of 3\,nW power broadening leads to a significantly increased value of 29\,MHz. We observe that for narrower linewidths, the photon number threshold required to reliably estimate the physical linewidth using the median is lower compared to broader linewidths. This is intuitive, since less photons are required to represent the line profile. 
These observations illustrate the limitations of standard data analysis techniques under noisy conditions and highlight the necessity for a more robust method to make a reliable prediction, here, enabling a valid optical characterization.


We develop a data analysis method based on Monte Carlo simulations that enables us to simulate the linewidth distribution statistics, as shown in Fig.~\ref{fig:exp}d, and derive robust parameter estimates even for severely undersampled data. We briefly outline the key steps of this method:
A single PLE scan is simulated by sampling the number of detection events $n$ from a normal distribution $\mathcal{N}(\bar{n}, \sigma)$, where $\bar{n}$ and $\sigma$ represent the fixed mean and standard deviation, respectively. These detection events are then distributed across a fixed frequency range of 150\,MHz according to a Cauchy distribution $P(\omega, \gamma) = \gamma / \pi(\omega^2 + \gamma^2)$, with a fixed linewidth $\gamma$. Similarly, the number of noise events is sampled from a Poisson distribution with a mean of two (see Supplemental Material for details). 
Following the experimental procedure, the resulting spectrum is fitted with a Voigt profile and the FWHM is recorded. This process is repeated multiple times.  
The resulting linewidths are compiled into a histogram for further analysis.
By comparing the simulated histograms with the experimental linewidth distribution, we can estimate the linewidth $\gamma$ of the Cauchy distribution, as well as the mean number of photon detection events $\bar{n}$. This is achieved through a $\chi^2$-test \cite{Avni1976}, where the function $S(\gamma, N)$ is minimized to find the best-fitting parameters (see Fig.~\ref{fig:MC_method}) with
\(
    S(\gamma, N) = \sum_{i = 0}^N [O_i - E_i(\gamma, N)]^2/E_i(\gamma, N)~,
\)
where, $O_i$ corresponds to the occurrences of the observed linewidths and $E_i(\gamma, N)$ to the occurrences of the simulated linewidths in a frequency bin $i$. 
The $99\%$ confidence interval can then be constructed \cite{Avni1976} by finding all $\gamma$ and $\bar{n}$ for which 
\(
    S(\gamma, \bar{n}) \leq S(\gamma_{\rm min}, \bar{n}_{\rm min}) + 9.21~. 
\)
Fig.~\ref{fig:exp}e shows the main result of our work: a linewidth prediction for $1$~nW and $3$~nW PLE scans that makes use of the MCM in a regime of $\bar{n} < 40$ where conventional statistical methods such as the median begin to become unreliable. For both powers the MCM predicts the true linewidth within the 99\% confidence intervals.   

\begin{figure}[h!]
    \centering
    \includegraphics[width = 0.75\columnwidth]{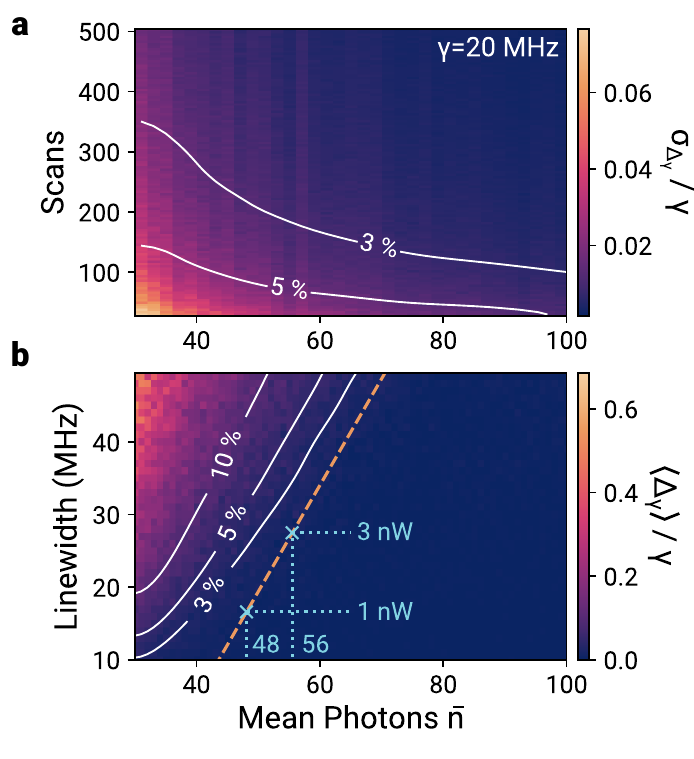}
    \caption{(a) Standard deviation $\sigma$ of the bias of the linewidth prediction of the median for $\gamma = 20$\,MHz as a function of the number of scans and the mean photon number $\bar{n}$. 
    (b) Bias of the linewidth prediction of the median as a function of the linewidth and $\bar{n}$ for a fixed number of scans $k=2000$. The orange dashed line indicates the threshold at which the estimated median linewidth has a 99\% probability of falling within a 2\% frequency window relative to the true value. The markers highlight the threshold photon numbers for the estimated linewidths of 17\,MHz and 29\,MHz at excitation powers of 1\,nW and 3\,nW (compare with Fig.~\ref{fig:exp}e).}
    \label{fig:median_stats}
\end{figure}

%
We employ the bias and consistency of a predictor, to evaluate and compare the performance of the median and the MCM. Here, we use the synthetic data generation technique with $\sigma = 6$ because it is close to the experimentally observed standard deviations for photon numbers $\bar{n} < 80$. We vary the linewidth $\gamma$ and mean photon number $\bar{n}$ for a total of $k=2000$ scans for each parameter set. The simulation is repeated 200 times to calculate averages. 
We evaluate the bias and consistency as a function of line scans and determine a threshold for a consistent and unbiased prediction. The absolute bias is
\(
\Delta_\gamma = |\langle \hat{\gamma}_k \rangle - \gamma|
\)
where $\langle\hat{\gamma}_k\rangle$ is the average linewidth predicted with the median after $k$ line scans, and $\gamma$ is the true linewidth. A good predictor shows no bias $\Delta_\gamma = 0$ for $k\rightarrow\infty$. 
Fig.~\ref{fig:median_stats}a shows the standard deviation $\sigma$ of the bias of the linewidth prediction for $\gamma = 20$ MHz as a function of the number of scans $k$ and the mean photon number $\bar{n}$. 
For this example, the median prediction's relative fluctuations stabilize to below 3\% for $k \gtrapprox 350$ scans. 
\begin{figure*}[ht!]
    \centering
    \includegraphics[width = 0.95\textwidth]{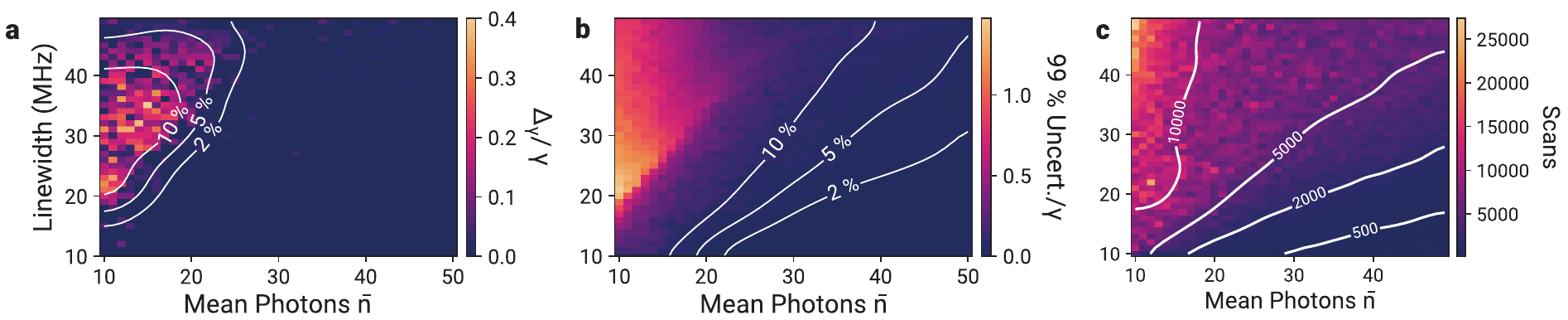}
    \caption{(a) Bias and (b) relative width of the 99\% confidence intervals of the Monte Carlo prediction as a function of linewidth and the mean photon number for $k=2000$ scans. (c) Number of scans for valid linewidth predicition with the MCM. The thresholds indicates where the 99\% confidence intervals are within 2\% of the true linewidth $\gamma$.}
    \label{fig:mc_combined}
\end{figure*}
The magnitude of the bias depends on both $\gamma$ and $\bar{n}$, as shown in Fig.~\ref{fig:median_stats}b. It is confirmed that broader linewidths require more photons to achieve an accurate prediction. When a resonance spans a greater number of frequency bins, a higher photon count is needed to adequately populate these bins, ensuring an accurate fit. The median shows a clear bias for $\bar{n} < 40$ quickly exceeding 5\% of the linewidth. The bias does not diminish with additional scans, which can be explained by the distributions observed in Fig.~\ref{fig:exp}d: a significant number of narrow linewidths skews the median prediction towards smaller linewidths.    

Further, evaluating the consistency of the median, aids in determining when it becomes a well behaved predictor. We use the definition of consistency \cite{bohm_introduction_2017}  
\\
\(
 {\rm Pr}\left(\lim_{k \rightarrow\infty}\hat{\gamma}_k = \gamma\right) = 1 ~,  
\) 
which demands that the probability of predicting the true linewidth approaches one for $k \rightarrow \infty$. 
In Fig.~\ref{fig:median_stats}b, the orange dashed line indicates the threshold where the estimated median linewidth has a 99\% probability of being within a 2\% frequency window relative to the exact value. We also use the consistency measure to quantify the precision of the median. The median becomes adequately precise once $\bar{n}$ surpasses the threshold for a given linewidth $\gamma$.
We found that in cases where the median exhibits significant bias, the confidence intervals (not shown here) based on the distribution of medians can become narrower than the bias itself as the number of scans increases. These narrow confidence intervals for $k \rightarrow \infty$ highlight an additional limitation of the median, indicating that it can yield a very precise but inaccurate prediction.
By setting thresholds for the desired accuracy and consistency, we can define a parameter range for $\gamma$ and $\bar{n}$ where the median is a straightforward predictor. 
Comparing synthetic data with experimental results shows that the consistency threshold for the median is reached at  $\bar{n} \approx 48$ for 1~nW and $\bar{n} \approx 56$ for 3~nW. These values correspond closely with the $\bar{n}$ where median values begin to stabilize close to the expected linewidth, as seen in Fig.~\ref{fig:exp}e (see~\cite{orphal-kobin_optically_2023} and Supplemental Material for comment on other standard methods).     


We now analyze the MCM in a parameter range where the median fails to be a good predictor. 
Fig.~\ref{fig:mc_combined}a shows the bias of the MCM as a function of the linewidth and mean photon number $\bar{n}$. The MCM exhibits a bias exceeding 2\% relative to the true linewidth for $\bar{n} < 30$ across a broad range of linewidths. Notably, the magnitude of this bias is significantly smaller compared to the median prediction within the same parameter range (compare with Fig.~\ref{fig:median_stats}b). In Fig.~\ref{fig:mc_combined}b, we present the 99\% confidence intervals of the Monte Carlo prediction, which are much wider than the bias observed in Fig.~\ref{fig:mc_combined}a, indicating much improved statistical behaviour compared to the median prediction. 
The MCM as presented here is always consistent, provided that the underlying model accurately represents the experimental system. 
We further determine the number of scans required to achieve the defined precision threshold. Fig.~\ref{fig:mc_combined}c shows the number of scans needed to meet this benchmark.

\begin{figure}[ht!]
    \centering
    \includegraphics[width = 0.7\columnwidth]{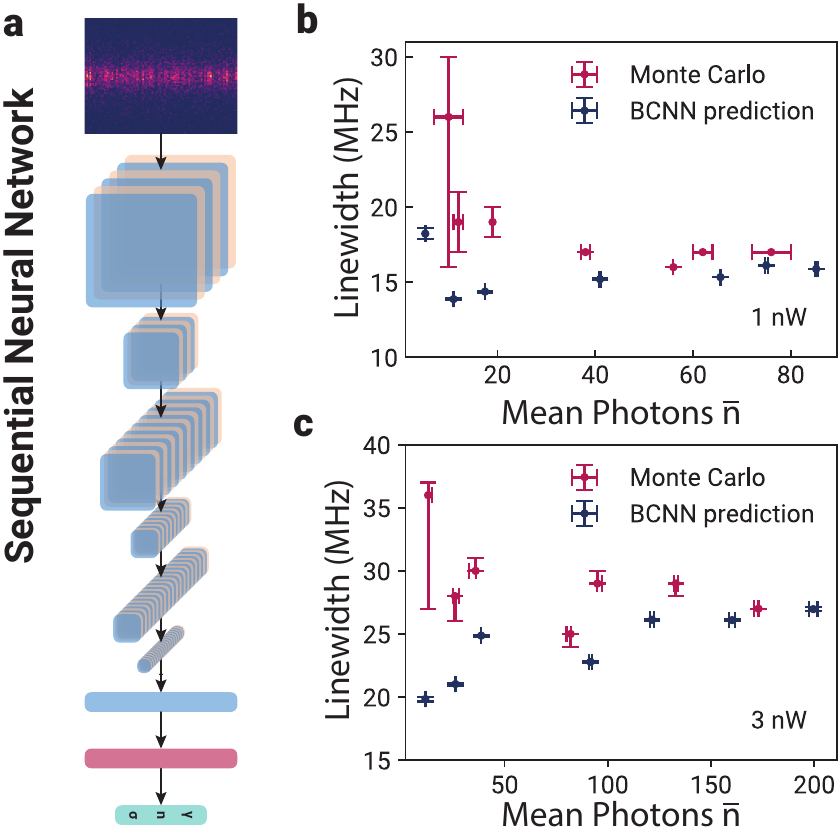}
    \caption{(a) BCNN network for prediction of the linewidth $\gamma$, the mean photon number $\bar{n}$ and the standard deviation $\sigma$ of the distribution of $n$. (b), (c) BCNN prediction of the experimental $1$\,nW data and $3$\,nW for both $\gamma$ and $\bar{n}$ ($\sigma$ not shown). The BCNN prediction shows good qualitative agreement with $\gamma$, but lacks predictive accuracy for $\bar{n}$.} 
    \label{fig:deep_learning}
\end{figure} 
%
%
We briefly explore deep learning for parameter prediction~\cite{borodinov_deep_2019, nguyen_deep_2021, nolan_machine_2021, xiao_parameter_2022, yang_entanglement_2024}. This technique can bypass the fitting procedure by training on unprocessed simulated line scans. 
Fig.~\ref{fig:deep_learning}a illustrates a LeNet-inspired model architecture~{\cite{lecun_gradient-based_1998} to estimate the parameters $\gamma$, $\bar{n}$, and $\sigma$. The model is a sequential Bayesian convolutional neural network (BCNN) with an input layer of dimension $75 \times 500$ representing 500 line scans across 75 frequency bins each(see Supplemental Material for details).    
A comparison of the 1\,nW and 3\,nW MCM predictions and the machine learning predictions is shown in Fig.~\ref{fig:deep_learning}b and c. The BCNN prediction shows promising agreement. We calculate uncertainties by sampling from the prior distribution. A notable deviation occurs in the predicted mean photon number for both data sets, which decreases for $\bar{n}<20$. The linewidth prediction shows a much smaller bias than the median prediction.


In conclusion, we introduced a method for information reconstruction from a set of undersampled data based on Monte Carlo simulations. This approach is illustrated by an experimental application example—estimating the optical linewidth of a quantum emitter from PLE data. Our results, supported by both experimental and synthetic data, show that the MCM can deliver highly accurate predictions in parameter regimes where conventional predictors in statistical data analysis, such as the median, fail to provide reliable results. According to our analysis, the median value can be used to achieve the target precision and accuracy of the linewidth estimation, if a minimum threshold for the mean photon number and number of scans has been reached. This figure of merit provides guidance for conducting valid data analysis in PLE experiments with low signal detection, even by applying simple standard methods.
As an outlook, although, the initial implementation of a BCNN is outperformed by the MCM, it is expected that an optimized routine based on deep learning will surpass in parameter prediction efficiency, for example, by using deeper convolutional networks~\cite{krizhevsky_imagenet_2017} and experimental training data.
Finally, we conclude that the MCM is broadly applicable to any situation where data on a physical constant become incomplete due to low signal transfer and subsequent statistical errors. This includes applications such as Raman spectroscopy, quantum state tomography, photon correlation measurements, and quantum interference experiments, unlocking new opportunities for investigations under challenging experimental conditions.


\textit{Acknowledgments--}The authors are grateful to the group of Ronald Hanson for providing the sample. The authors acknowledge funding by the European Research Council (ERC, Starting Grant project QUREP, No. 851810) and the German Federal Ministry of Education and Research (BMBF, project QPIS, No. 16KISQ032K; project QR.X, No. KIS6QK4001).

\textit{Data availability--}The data that support the findings of this study are available from the corresponding author upon reasonable request.

\bibliography{linewidth.bib}

\end{document}



\title{Supplemental Material: Retrieving Lost Atomic Information: Monte Carlo-based Parameter Reconstruction of an Optical Quantum System}

\author{Laura Orphal-Kobin}\thanks{L.O.-K. and G.P. contributed equally to this work.}
\affiliation{\vspace{0.5em}Department of Physics \& IRIS Adlershof, Humboldt-Universität zu Berlin, Newtonstr. 15, 12489 Berlin, Germany}
\author{Gregor Pieplow}\thanks{L.O.-K. and G.P. contributed equally to this work.}
\affiliation{\vspace{0.5em}Department of Physics \& IRIS Adlershof, Humboldt-Universität zu Berlin, Newtonstr. 15, 12489 Berlin, Germany}
\author{Alok Gokhale} 
\affiliation{\vspace{0.5em}Department of Physics \& IRIS Adlershof, Humboldt-Universität zu Berlin, Newtonstr. 15, 12489 Berlin, Germany}
\author{Kilian Unterguggenberger} 
\affiliation{\vspace{0.5em}Department of Physics \& IRIS Adlershof, Humboldt-Universität zu Berlin, Newtonstr. 15, 12489 Berlin, Germany}
\author{Tim Schröder}
\email{tim.schroeder@physik.hu-berlin.de}
\affiliation{\vspace{0.5em}Department of Physics \& IRIS Adlershof, Humboldt-Universität zu Berlin, Newtonstr. 15, 12489 Berlin, Germany}
\affiliation{\vspace{0.5em}Ferdinand-Braun-Institut, Gustav-Kirchhoff-Str. 4, 12489 Berlin, Germany}

\maketitle

\renewcommand \thesection {S\arabic{section}} 
\setcounter{figure}{0}
\renewcommand{\figurename}{Fig.}
\renewcommand{\thefigure}{S\arabic{figure}}
\renewcommand \thetable {S\arabic{table}} 
\renewcommand \theequation {S\arabic{equation}} 

\section*{Experiment}
\subsection{Sample}
A single natural low-strain NV in a type IIa chemical vapor deposition grown diamond (Element 6) is used. To enable a larger photon collection efficiency, a solid immersion lens is deterministically etched around the emitter. More details on the sample and fabrication can be found elsewhere~\cite{robledo_high-fidelity_2011}. We assume that the investigated NV exhibits negligible spectral diffusion on the time scale of a single PLE scan and ``single-scan'' linewidths are predominantly power broadened.

\subsection{Experimental Details}

The experiments are performed in a home-built confocal microscopy setup at cryogenic temperatures at about 4\,K in a closed-cycle helium cryostat (AttoDRY800). The signal photons are collected through an objective with a numerical aperture of 0.82, coupled into a single-mode fiber, and are detected by two avalanche photodiodes (SPCM-AQRH, Excelitas Technologies). The transmission in the detection path is adjusted by a continuously variable metallic neutral density filter. The set point of the transmission is selected by recording the photon counts at continuous excitation by a diode laser at 525\,nm (DLnsec, LABS GmbH). 

In PLE measurements, the frequency of a tunable diode laser (New Focus Velocity TLB-6704) is tuned by applying an external piezo voltage at a scan speed of 6\,GHz/s. For every parameter setting 3200 line scans are recorded while the laser frequency is tuned back and forth~\cite{binder_qudi_2017}. For initialization of the NV in the negative charge state, a laser pulse at 525\,nm with a duration of 10\,ms and a power of 2.3\,mW is applied before every line scan. 
Additionally, a second laser (Toptica DL pro HP) at 637\,nm resonant to the optical $\ket{m_{\rm s}=\pm 1} \Leftrightarrow \ket{A_1}$ transitions (cw, 20 nW) is applied, resulting in a permanent optical spin pumping into the $\ket{m_{\rm s}= 0}$ spin state. In this way, a bright and stable signal can be maintained. The scanning frequency of the tunable diode laser is recorded by a wavelength meter (HighFinesse WS7) at an integration time of 2\,ms, and moreover, the wavelength of the second laser is stabilized.

Lifetime measurements are performed using a supercontinuum white light laser (NKT Photonics SuperK FIANIUM FIU-15) at a central wavelength of 532\,nm, a bandwidth of 10 nm, and a repetition rate of 1.2\,MHz at a cw excitation power of 20\,µW. Second-order correlation measurements are performed in a Hanbury Brown–Twiss setup configuration at 525\,nm cw laser excitation. Photon detection events are recorded and cross-correlated using the time-tagger quTAG MC by quTools.

\subsection{Linewidth Analysis} 
In the data analysis, the bin width is set to 2.2\,MHz/step. Data sets are considered for linewidth estimation when the signal-to-noise ratio is at least 3:1 in at least one bin, meaning that a minimum of 3 photons must be collected in an individual frequency bin.
The resonance in a single PLE scan is fitted with a Voigt profile by a least-squares minimization algorithm in the implementation provided by the lmfit Python package~\cite{newville_lmfit_2014}. The Voigt distribution function is modelled by
\begin{align}
    f(x;A,\mu,\sigma,\gamma) = \frac{A \textrm{Re}[w(z)]}{\sigma\sqrt{2\pi}} 
\end{align}
where 
\begin{align}
   z = \frac{x-\mu+i\gamma}{\sigma\sqrt{2}} 
\end{align}
and
\begin{align}
   w(z) = e^{-z^{2}}\textrm{erfc}(-iz) ~.
\end{align}
Here, $A$ is the amplitude and $\mu$ is the center. The FWHM is approximately 3.6013$\sigma$.

\subsection{Quantum Emitter Characterization} 

\begin{figure*}
\centering
\includegraphics[width=1.0\textwidth]{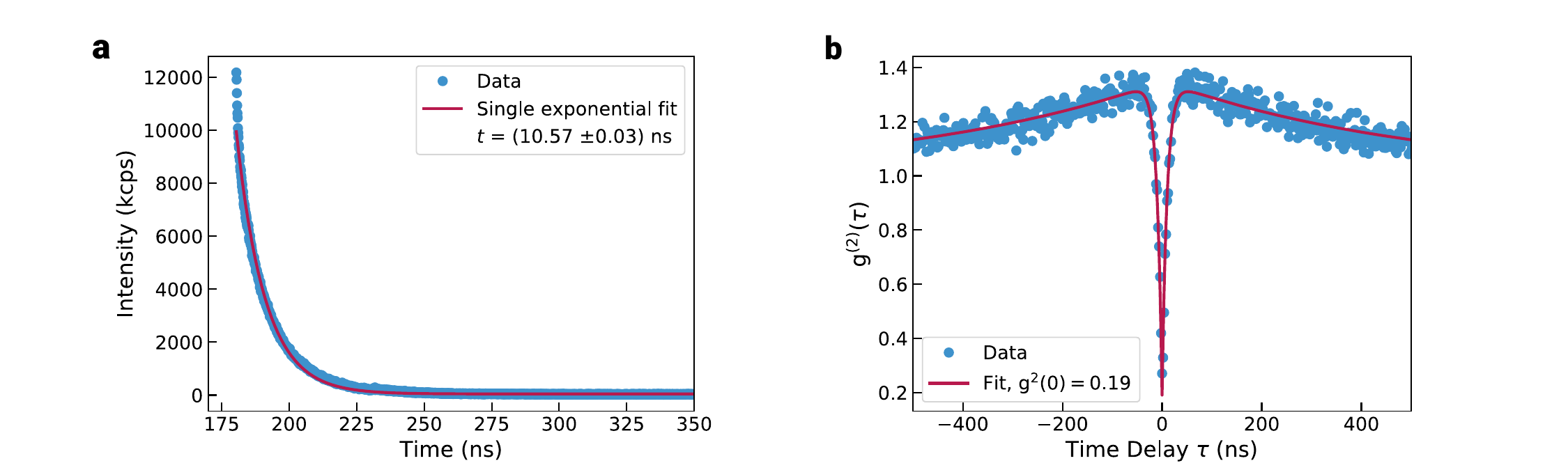}
\caption{\label{fig:LTg2} Quantum emitter characterization. (a) The lifetime of the NV is extracted by a single exponential fit of a lifetime measurement at 532\,nm pulsed excitation.  (b) In a second-order autocorrelation measurement with cw excitation at 525\,nm ($P=1.8$\,mW), antibunching is observed, proving that a single quantum emitter is investigated.}
\end{figure*}

From a lifetime measurement, the lifetime of the NV is determined by applying a single exponential fit (see Fig.~\ref{fig:LTg2}a). The lifetime-limited linewidth sets the physical lower bond of the quantum emitter's linewidth ($\tau = 1/{(2\pi\Delta\nu)}$). Here, the NV's lifetime is 10.57\,ns corresponding to a natural linewidth of 15.1\,MHz.

In a second-order autocorrelation measurement, antibunching is observed (see Fig.~\ref{fig:LTg2}b). A dip with a minimum of 0.19 ($<0.5$) at zero time delay indicates that a single quantum emitter is investigated. The NV is approximated as a three-level system and the correlation measurement can be described by the fit function
\begin{equation}
    g^{2}(\tau) = 1 + p_f^2\left(\underbrace{c\,e^{-\vert\tau\vert/\tau_b}}_{\text{bunching}} - \underbrace{(1+c)\, e^{-\vert\tau\vert/\tau_a}}_{\text{antibunching}}\right) ~.
\end{equation}

\section{Comparison of standard methods for data analysis.} 
\begin{figure*}[ht]
\centering
\includegraphics[width=0.95\textwidth]{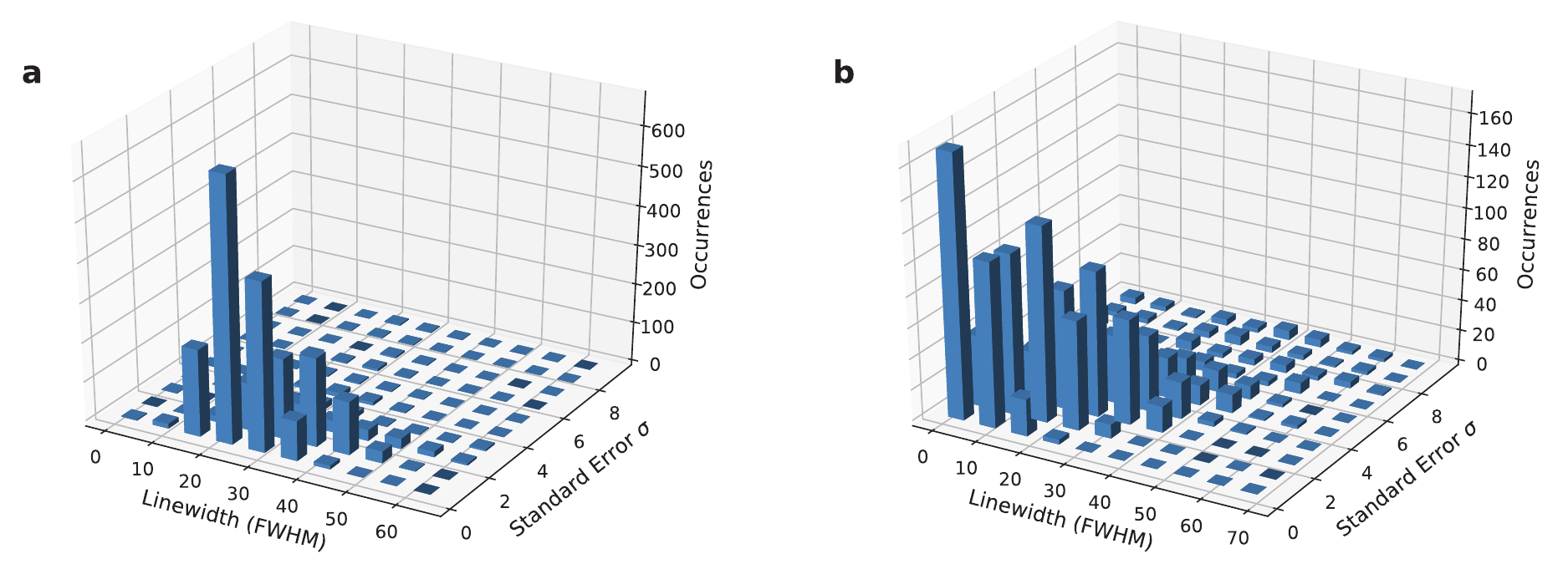}
\caption{\label{fig:err_correl} Histogram summarizing the occurrences of recorded linewidths and standard errors obtained from the profile fits. Transmission in detection path of (a) 100$\%$ corresponding to a mean photon number of 176 and (b) transmission of 10$\%$ corresponding to a mean photon number of 25 in a frequency window of 150 MHz around the resonance.}
\end{figure*}

\begin{figure*}[ht]
\centering
\includegraphics[width=0.95\textwidth]{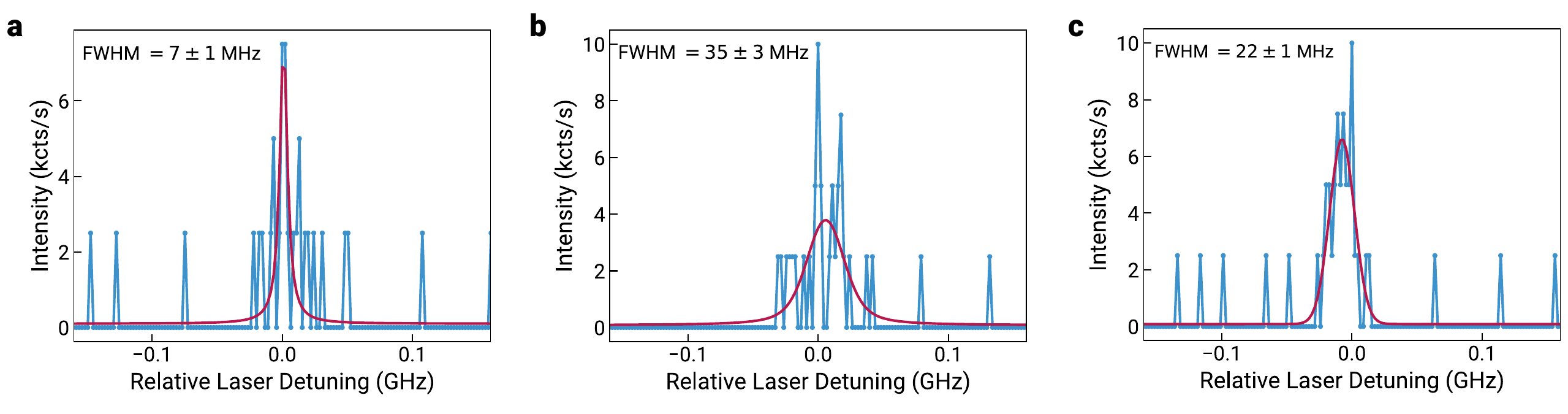}
\caption{\label{fig:SingleLines} Examples for single PLE scans with low photon number $\bar{n}=25$ at signal transmission of 10\% at an excitation power of 3\,nW. When the signal is only slightly above the noise level, standard fitting routines fail to accurately extract the physical linewidth. In (a), the intensity peak is fitted as a narrow resonance with low uncertainty and (b) shows a multi-photon detection event fitted as broad profile with large uncertainty. (c) Shows a reasonable profile sampling and fit of the ``single-scan'' linewidth.}
\end{figure*}

In the quantum optics community, particularly in studies involving color centers in diamond, inverse variance weighting and the median value are common methods for estimating the homogeneous linewidth of quantum emitters~\cite{yurgens_spectrally_2022,chakravarthi_impact_2021,ruf_optically_2019,van_dam_optical_2019}. 
Inverse variance weighting (IVW) is a standard statistical method that assigns more weight to estimates with higher precision (i.e., smaller uncertainties), under the assumption that the uncertainties are uncorrelated with the estimates themselves.
However, this assumption can break down in practice, especially in cases where the data is undersampled. Figure~\ref{fig:err_correl} illustrates this issue by showing histograms of estimated linewidths extracted from experimental data alongside their corresponding standard errors. It becomes apparent that, particularly for low photon detection efficiencies, there is a correlation between the estimated linewidths and their (absolute) uncertainties: larger linewidths are associated with larger uncertainties, while narrower linewidths tend to have smaller uncertainties.
This phenomenon can be easily explained: When photon detection events above the noise level are sparse, the fitting routine may either identify a narrow linewidth with little or no discrepancy or fit a broader linewidth with greater discrepancies (see example in Fig.~\ref{fig:SingleLines}). As a result, narrow linewidths are often extracted with seemingly higher precision, even though this precision is illusory due to the low signal strength.
The key consequence of this effect is that the average linewidth estimated using IVW is biased towards narrower values, leading to an underestimation. 
To address this issue, one could consider redefining the filter settings to exclude PLE scans with insufficient photon counts above the noise level. However, this solution comes with the trade-off of potentially requiring much longer measurement times to achieve statistically significant results. 

The median, which represents the middle value of a sorted data set, is a robust alternative for linewidth estimation. It is less sensitive to outliers and skewed data distributions, and not vulnerable to correlated errors. Moreover, the median of a data set can be determined easily. For the investigated data sets, the median of the fitted linewidths and the median extracted from a lognormal distribution fit were in agreement. 

\section{Statistical Methods}

\subsection{Details on Monte Carlo linewidth simulation} 

We reject lines, if no frequency bin (2\,MHz width) with at least three photon counts is present. 

Details on binning: Even though the simulations qualitatively reproduce a large occurrence of unphysically narrow lines they still don't match the observed data perfectly. Smaller linewidths $\gamma \leq 15 \rm\,MHz$ are therefore counted in a single bin. The binning for larger linewidths has to be adjusted according to both, the observed and simulated data because frequency bins with fewer than $5$ counts on average will make the fitting procedure more unreliable. The details of the procedure can be found in \cite{Avni1976}.

\subsection{Comment on Photon Statistics and Number of Photons} 
\begin{figure*}[ht]
\centering
\includegraphics[width=1.0\textwidth]{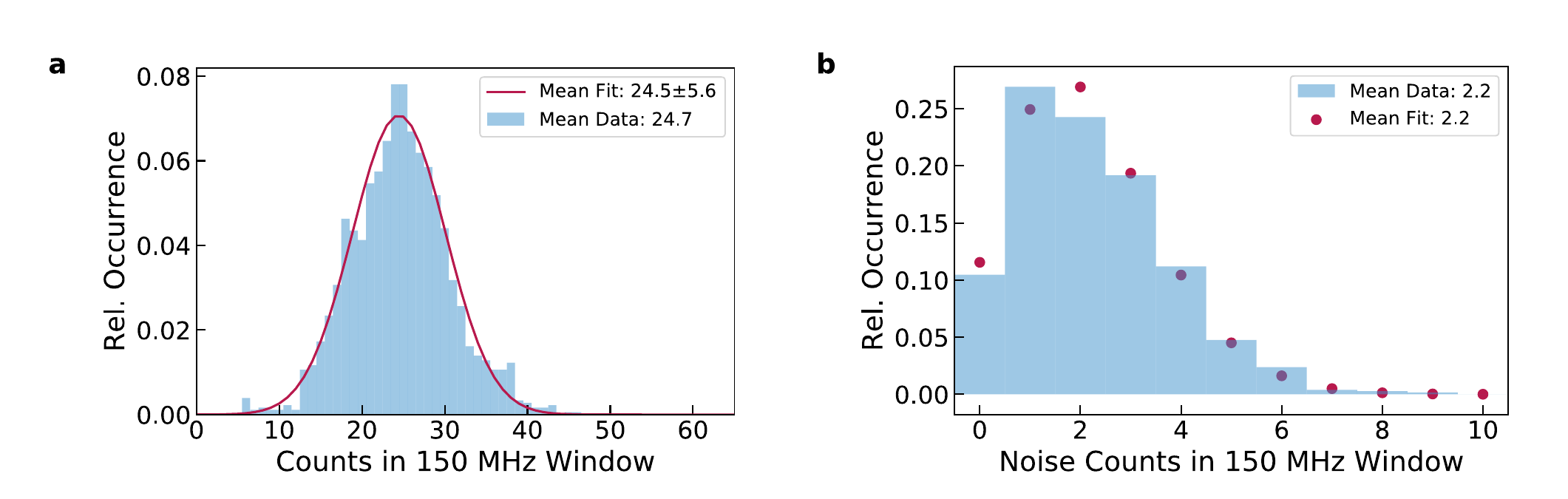}
\caption{\label{fig:PhoStat} Photon statistics of detected photons. (a) The signal photons are counted in a frequency window of 150\,MHz around the resonance. Here, exemplary the 3\,nW data for 10\% signal transmission are shown. The distribution of photon numbers is described by a normal distribution function. (b) The noise photons are counted in a 150\,MHz window 500\,MHz apart from the resonance. The photon statistics show Poisson behavior. }
\end{figure*}

In photon counting experiments with a single quantum emitter, one typically expects a sub-Poisson distribution of the detected photon counts. However, photon statistics are highly sensitive to losses and inefficient detection, which can degrade the expected statistics. In this case, and under the assumption of random detection processes, Poisson statistics apply, which is commonly used to describe coherent light sources~\cite{fox_quantum_2006}. In experimental data analysis, the mean number of photons in a 150 MHz range window is characterized by a normal distribution which approximates a Poisson distribution for large sample size, as shown in Fig.~\ref{fig:PhoStat}a. The number of background detection events is characterized by a Poisson distribution considering a frequency window 500\,MHz apart the resonance Fig.~\ref{fig:PhoStat}b.

\subsection{Number of PLE Scans} 
In agreement with Fig.\,4c in the main text, a more extensive linewidth analysis from experiment proves robustness against variations in the number of scans, provided that a sufficient number of photons are detected. For the case of full signal in the detection path, the estimated linewidth is compared taking 3200, 1200, 1000...100, 50 scans into account. The linewidth estimated using the median remains within a range where the difference between the minimum and maximum values is less than 10$\%$. Although the IVW method possibly tends to underestimate the linewidth, the variability of the estimates is comparable to that of the median approach.

\subsection{Machine Learning} 

The model we use for predicting the the linewidth, mean photon number and its variance is structured as follows: The input layer is a two-dimensional array with dimensions of $(500, 75)$, representing a total of 500 linescans over a 150 MHz window, with each bin covering $2$ MHz. 
The input layer is followed by two 32$\times$32 2D convolutional layers, each with a 2x2 kernel and a 2$\times$2 pooling layer. Next is a 128$\times$128 2D convolutional layer with a 3$\times$3 kernel, a flattening step, and a 512-neuron dense layer. Finally, a dense variational layer, assuming normal prior and posterior distributions, estimates the three parameters we intend to predict. The variational layer is used to estimate the uncertainty in the prediction. All convolutional and Dense layers had ReLU (Rectified Linear Unit) activation. We implement the network using the Python TensorFlow library~\cite{tensorflow2015-whitepaper}. 

We train the network using a set of 50000 spectra each generated for a random selection of $\gamma, \bar{n}$ and $\sigma$. For each epoch the training data is regenerated to avoid overfitting. We trained the model for 100 epochs with a mean squared error loss function. We performed the training on an RTX 3090 graphics card.

\begin{figure}
    \centering
    \includegraphics[width=0.45\columnwidth]{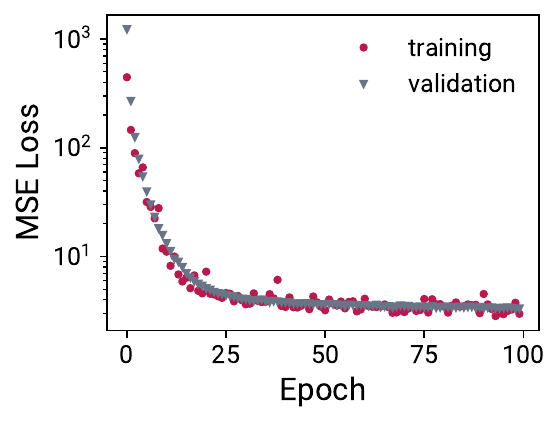}
    \caption{Mean squared error loss as a function of the training epochs of the BCNN.}
    \label{fig:ml_loss}
\end{figure}

Fig.~\ref{fig:ml_loss} shows the validation and training loss (mean squared error).

\begin{figure*}[ht!]
\centering
\includegraphics[width=1.0\textwidth]{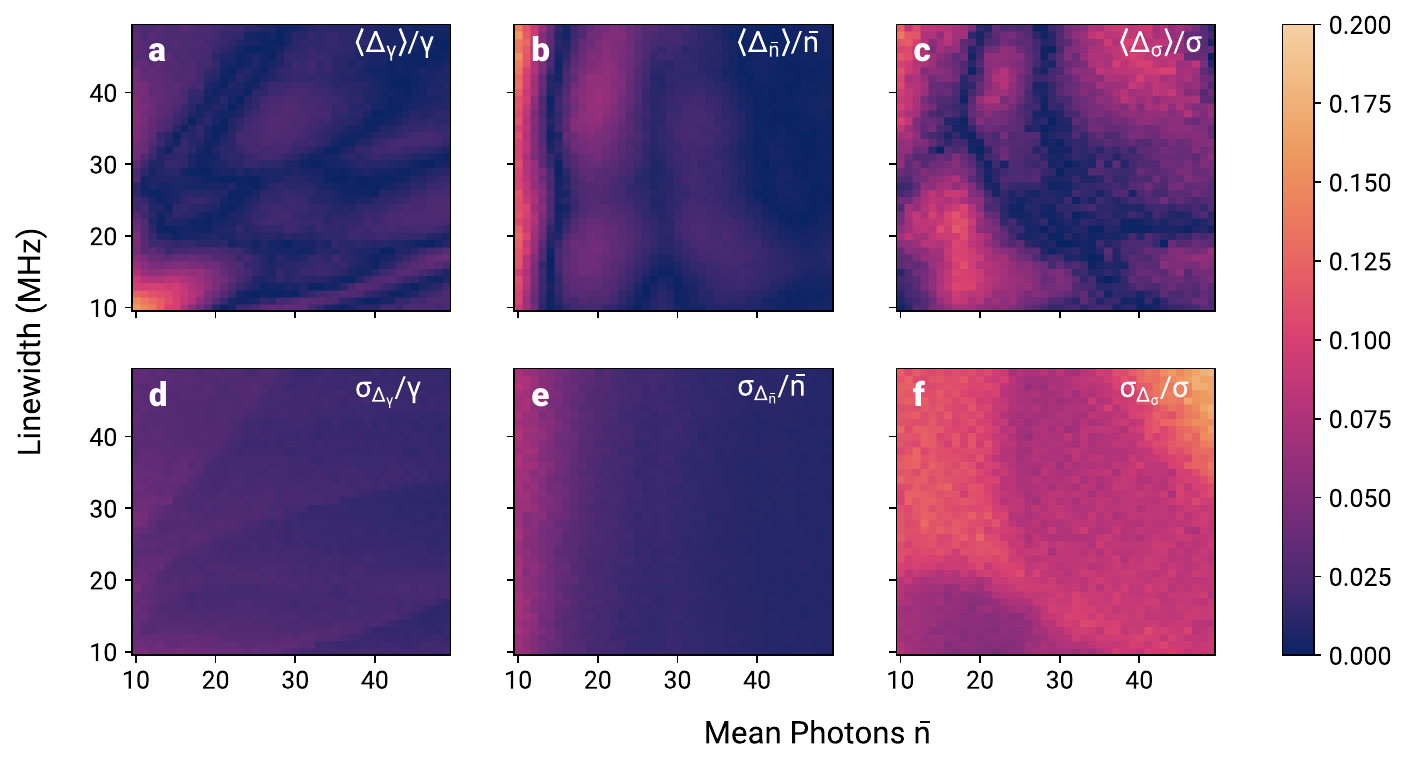}
\caption{\label{fig:combined_ml_var} 
Overview of the bias and stadard deviation of the BCNN prediction. Averages were performed for 200 predictions. A standard deviation of the mean photon number $\sigma = 6$ was chosen for the synthetic data. 
(a) Average relative bias $\langle \Delta_\gamma \rangle /\gamma$ of the BCNN prediction of $\gamma$. 
(b) Average relative bias $\langle \Delta_{\bar{n}} \rangle /\bar{n}$ of the BCNN prediction of $\bar{n}$.
(c) Average relative bias $\langle \Delta_{\bar{\sigma}} \rangle /\sigma$ of the BCNN prediction of $\sigma$. 
(d) Relative standard deviation $\sigma_{\Delta_\gamma} /\gamma$ of the BCNN prediction of $\gamma$. 
(e) Relative standard deviation $\sigma_{\Delta_{\bar{n}}}/\bar{n}$ of the BCNN prediction of $\bar{n}$. 
(f) Relative standard deviation $\sigma_{\Delta_\sigma} /\sigma$ of the BCNN prediction of $\sigma$.
}
\end{figure*}

Fig.~\ref{fig:combined_ml_var} shows the biases and standard deviations for the parameters that we predicted with the BCNN. We used synthetic data for the prediction. A significant relative bias can be seen for both $\gamma$ and $\bar{n}$ in Fig.~\ref{fig:combined_ml_var}a,b for small photon numbers and small linewidths. The bias and standard deviation in the prediction of $\sigma$ is much more significant for the prediction of $\sigma$, showing that the BCNN does not perform well for this parameter. The bias in $\bar{n}$ is noticably small for $\bar{n} > 40$, indicating, that the synthetic data performs much better than the experimental data. We believe this is due to the stronger fluctuations in mean number of photons in the experimental data, which is not well predicted by the BCNN (see Fig. \ref{fig:combined_ml_var}c,f.). The weak performance of the BCNN is probably caused by the type of LeNet network architecture, which is typically used for image recognition.  

\bibliography{linewidth.bib}